\documentclass[useAMS,usenatbib,asm]{mn2e}
\usepackage{graphicx,fleqn,times}
\usepackage{graphicx}
\usepackage{color}

\def\today{\ifcase\month\or
 January\or February\or March\or April\or May\or June\or
 July\or August\or September\or October\or November\or
 December\fi\space\number\day, \number\year}

\def\todmy{\number\day\space\ifcase\month\or
 January\or February\or March\or April\or May\or June\or
 July\or August\or September\or October\or November\or
 December\fi\space\number\year}

\newcommand{\tal}{\it et al. \rm}

\title{A bar in the inner halo of barred galaxies I. Structure and
  kinematics of a representative model}
\author[E. Athanassoula]
       {
       E. Athanassoula \\
Laboratoire d'Astrophysique de Marseille, Observatoire Astronomique de
  Marseille Provence, 2 Place Le Verrier, 
F-13248 Marseille Cedex 4, France \\
}

\date{Accepted .
      Received ;
      }

\pagerange{\pageref{firstpage}--\pageref{lastpage}}
\pubyear{2006}

\begin{document}

\maketitle

\label{firstpage} 
\begin{abstract}
$N$-body simulations argue that the inner haloes of barred galaxies
should not be spherical, nor even axisymmetric, but triaxial. 
The departure from sphericity is strongest near the
centre and decreases outwards; typical axial ratios for
the innermost parts are of the order of 0.8. The halo shape is
prolate-like in the inner parts up to a certain radius and then turns
to oblate-like. I call 
this inner halo structure the `halo bar' and analyse here in depth its
structure and kinematics in a representative model. It 
is always considerably shorter than the disc bar. It lags the disc bar
by only a few degrees at all radii and the difference between the two
bar phases increases with distance from the centre. The two bars turn with 
roughly the same pattern speed. This means that the halo bar is a
slow bar, since its corotation radius is much larger than its length. The
bisymmetric component in the halo continues well outside the halo bar
in the form of an open spiral, trailing behind
the disc bar. The inner parts of the halo display some mean rotation
in the same sense as the disc rotation. This is more important
for particles nearer to the equatorial plane and decreases with increasing
distance from it, but is always much smaller than the disc rotation.  
\end{abstract}


\section{Introduction}
\indent

The exchange of angular momentum between different parts of a barred
galaxy is of major importance for its
evolution. \citet{Lynden-BellKalnajs} first discussed this process in
the case of a spiral galaxy. Their paper was
followed by a number of others, discussing and establishing the
role of angular momentum exchange for various aspects of disc galaxy
evolution (\citeauthor{Mark} \citeyear{Mark}; \citeauthor{Kormendy}
\citeyear{Kormendy}; \citeauthor{Sellwood80} \citeyear{Sellwood80};
\citeauthor{TremaineWeinberg84} \citeyear{TremaineWeinberg84};
\citeauthor{Weinberg} \citeyear{Weinberg}; \citeauthor{TSAP}
\citeyear{TSAP}; \citeauthor{STAP} 
\citeyear{STAP}; \citeauthor{LittleCarlberga}
\citeyear{LittleCarlberga, LittleCarlbergb};
\citeauthor{HernquistWeinberg} \citeyear{HernquistWeinberg};
\citeauthor{Ath96} \citeyear{Ath96}; \citeauthor{DebattistaSellwood}
\citeyear{DebattistaSellwood}; \citeauthor{Ath02} \citeyear{Ath02},
hereafter A02; \citeauthor{WeinbergKatz} \citeyear{WeinbergKatz};
\citeauthor{Ath03} \citeyear{Ath03}, hereafter A03;  
\citeauthor{ValenzuelaKlypin03} \citeyear{ValenzuelaKlypin03}; 
\citeauthor{Fuchs} \citeyear{Fuchs}; \citeauthor{Ath05b}
\citeyear{Ath05b}; \citeauthor{FuchsAthanassoula}
\citeyear{FuchsAthanassoula}; \citeauthor*{HolleyBockelmann+05}
\citeyear{HolleyBockelmann+05}; \citeauthor{Debattista+06} 
\citeyear{Debattista+06}; \citeauthor*{MartShlo2}
\citeyear{MartShlo2}). Most of these studies are based on $N$-body simulations.

\citet{Lynden-BellKalnajs} showed that angular momentum is
transferred from the inner to the outer parts of the disc and that it is
mainly (near)-resonant material that will emit, or absorb
it. Material at inner Lindblad resonance (hereafter ILR) will lose
angular momentum, while material at corotation (hereafter CR) and outer
Lindblad resonance (hereafter OLR) will gain it. Spiral structure within 
CR is a disturbance with negative angular momentum, so
that feeding it with angular momentum will damp it, while taking
angular momentum from it will excite it \citep[see also][]{Kalnajs}.

This insightful analytical  picture did not include any spheroidal
component. It was thus extended 
later in the context of bar slow down by \citet{TremaineWeinberg84},
\citet{Weinberg} and Hernquist \& Weinberg (1992), and in the context
of bar growth by A02 and A03. The halo has a different behaviour
from that of the disc, since, provided the halo distribution function
is a function of the energy only, it is possible to show that halo
particles at all resonances will gain angular momentum. It should be
possible to extend this property to a more general class of distribution
functions, provided energy is the main functional
dependence and an appropriate perturbation expansion can be used. If
both the outer disc and the halo can absorb angular 
momentum, then the inner disc can emit more than it would in the
absence of a halo, and this, since the bar is a negative angular
momentum `perturbation', will lead to a stronger bar than if only the
outer disc is absorbing (A02; A03). This means that 
stronger bars can be found in models whose haloes have higher
densities in the resonance regions, and thus
explains the results found by Athanassoula \& Misiriotis (2002,
hereafter AM02). 

As the bar loses angular momentum it will grow stronger (A02; A03) 
and will also slow down, i.e. its pattern
speed will decrease (Weinberg 1985; Little \& Carlberg
1991a,b; Hernquist \& Weinberg 1992; Athanassoula 1996; Debattista \&
Sellwood 2000; A03; O'Neill \& Dubinski 2003; Martinez-Valpuesta \tal 2006). 
The amount of angular momentum gained/lost at a given resonance
depends on the density there, but also on how cold the material at
resonance is (A03), since colder material can emit/absorb much more angular
momentum than hot material. Thus, these two factors determine the
angular momentum exchange and therefore the morphology, strength and
angular velocity of the bar (A03; Athanassoula 2005b).

Although a number of studies have addressed the effect of the bar
evolution on the disc component (e.g. Debattista \& Sellwood 2000;
AM02; A02; A03; O'Neill \& Dubinski 2003; Valenzuela \& Klypin
2003; Martinez-Valpuesta \& Shlosman 2004; Holley-Bockelmann \tal
2005; Debattista \tal 2006; 
Martinez-Valpuesta, Shlosman \& Heller 2006), very  
few have focused on the effect on the halo component. Most of these 
address the effect of the bar on the halo radial density profile
(Hernquist \& Weinberg 1992; Weinberg \& Katz 2002; Sellwood 2003;
Athanassoula 2004; Holley-Bockelmann \tal 2005; McMillan \& Dehnen
2005; Sellwood 2006), 
while a few address other aspects of the halo evolution. Debattista \& 
Sellwood (2000) mentioned an induced bisymmetric distortion in the halo.
A02 showed that a considerable fraction of the halo particles is at
resonance with the bar. In particular at CR, but also at ILR,
OLR and at several other resonances. This was further confirmed in
A03, where it was also shown that, as predicted analytically for
distribution functions that depend on the energy only, 
the halo resonances absorb angular momentum. Further confirmation,
for different models and/or with different techniques have been also
given by Ceverino \& Klypin (2005) and by Martinez-Valpuesta \tal
(2006). O'Neill \& 
Dubinski (2003) also discussed the existence of a bar in the halo
component and mentioned an axis ratio of 0.88. A bar in
the halo component is shown in fig. 2 of Holley-Bockelmann \tal
(2003).  
Various aspects of the effect of a non-axisymmetric halo on bar
evolution were discussed by El-Zant \& Shlosman (2002), Berentzen,
Shlosman \& Jogee (2006), Berentzen \& Shlosman (2006) and Heller,
Shlosman \& Athanassoula (2007).

Three works discussed properties of the halo bar in
some detail. Hernquist \& Weinberg (1992) were the first to discuss
the halo bar and gave axial ratio values of 0.7 and 0.8-0.9 for $c/a$ and 
$b/a$, respectively. They also mentioned that the halo bar lags the
disc bar somewhat. CPU limitations at the time, however, made them use
a live halo but a rigid bar and no disc. Thus, their results do not
include the effect of the bar evolution on the halo
properties. As mentioned by the authors, this work needs to be
extended to cover a broader part of the parameter space and,
particularly, to include a live disc. 
Athanassoula (2004; 2005b; 2005c) discussed briefly some halo bar properties,
including the increase of the bar length with time. The third
of these papers discusses also the orbital structure in the halo,
including the chaos versus regularity question and the properties of
the near-resonant orbits. Finally, Colin, Valenzuela \& Klypin
(2006) discuss the properties of the halo bar, which they call `dark
matter bar', in cosmologically motivated simulations and find
properties in agreement with those in Athanassoula (2004; 2005b; 2005c). In
particular, they give an axial ratio of 0.7.

Here I will extend previous studies and discuss in depth the
properties of the halo bar in a 
number of high resolution, fully self-consistent $N$-body simulations. 
This paper is the first of a set, where I discuss the effects
of the disc bar and of its evolution on the halo component. In
$\S$~\ref{sec:simul}, I briefly summarise some relevant information
on all the simulations, both the one discussed here and those
discussed in the second paper of this series (Paper II). In
particular, I describe the different types 
of initial conditions used. In $\S$~\ref{sec:existence}, I introduce the
halo bar. Its shape and length are discussed in
$\S$~\ref{sec:inert} and \ref{sec:ellipse}. In $\S$~\ref{sec:length}, I
introduce spherical harmonics to measure the halo bar properties and
in particular its strength, length and position angle. 
Kinematics are discussed in $\S$~\ref{sec:kinematics} and I briefly
summarise in $\S$~\ref{sec:summary}. A discussion of the
implications of the results of this paper will be given in Paper II,
after I discuss the time evolution of the halo
bar properties and present statistics from a large sample of several
hundred simulations. 

\section{Simulations}
\label{sec:simul}
\indent

In this work, I consider simulations with a large variety of initial
conditions. In order to allow comparisons, however, all models share a
few common features. In particular, all
the models are composed of  a disc, a halo, and sometimes also a bulge
component. The disc has an initial volume density profile

\begin{equation}
\rho_d (R, z) = \frac {M_d}{4 \pi R_d^2 z_0}~~\exp (- R/R_d)~~{\rm sech}^2 (z/z_0),
\label{eq:discdens}
\end{equation}

\noindent
where $R$ is the cylindrical radius, $M_d$ is the disc mass and $R_d$ and
$z_0$ are the disc radial and vertical scale-lengths,
respectively. In all cases  $M_d$ = 1 and $R_d$ = 1. This
allows me to compare the results of the various simulations directly, without
any rescaling. I consider five families of simulations, with
different initial conditions for the spheroid.

The first family has the halo and bulge radial profiles described in
Hernquist (1993) and used in AM02, A02 and
A03 and is called here for brevity the AM family. The initial halo
density profile is   

\begin{equation}
\rho_h (r) = \frac {M_h}{2\pi^{3/2}}~~ \frac{\alpha}{r_c} ~~\frac {\exp(-r^2/r_c^
2)}{r^2+\gamma^2},
\label{eq:halodens}
\end{equation}

\noindent
where $r$ is the spherical radius, $M_h$ is the halo mass and $\gamma$
and $r_c$ 
are the halo scale-lengths. $\gamma$ can be considered as the core
radius of the halo. The constant $\alpha$ is defined by
$$
\alpha = [1 - \sqrt \pi~q~\exp (q^2)~~(1 - {\rm erf} (q))]^{-1},
$$
\noindent
where $q=\gamma / r_c$ (Hernquist 1993). 
If a particularly extended halo is necessary, the halo radial
density profile is described by the sum of two densities of the kind
given by Eq.~\ref{eq:halodens}. 

As shown in AM02 and as explained in A02 and A03, the size of the halo
core strongly influences 
the bar evolution. Haloes with a small core have a lot of mass in
the inner regions and thus, provided their velocity dispersion 
is not too high, they can contribute substantial angular
momentum sinks and lead to considerable angular momentum
exchange between the near-resonant particles in the bar region and
the near-resonant particles in the halo. Such models 
grow strong bars, i.e. bars that
are long, thin and massive and have rectangular-like
isodensities (AM02). Viewed side-on, i.e. edge-on with the
line-of-sight along the bar minor axis, they exhibit a strong peanut, or
even X-like shape. Such models are termed MH in the above mentioned works. On
the contrary, haloes with large cores have considerably less 
material in the inner parts and lead to less angular momentum
exchange. They are termed in the above mentioned works as MD. 
Bars grown in such environments are less strong, have 
elliptical-like isodensities when viewed face-on and boxy-like when
viewed side-on. More on the properties of these bars,
their different 
evolution and the explanation of these differences can be found in
AM02, A02, A03 and Athanassoula (2005a; 2005b). Let me stress, however,
that there is no discontinuity between the two types of models. On the
contrary, MH and MD types are the 
two extremes of a continuous sequence, based on a continuous distribution of
core sizes and of amount of angular momentum exchanged.

The AM models sometimes have also a bulge
component. Whenever present, this has an initial density profile

\begin{equation}
\rho_b (r) = \frac {M_b}{2\pi a_b^2}~~\frac {1}{r(1+r/a_b)^3},
\label{eq:hernquistp}
\end{equation}

\noindent
where $M_b$ is the bulge mass and $a_b$ is the bulge scale-length. I will
hereafter refer to this as the Hernquist profile (Hernquist 1990). 

The adding of the disc and spheroidal components together is not
trivial and I will describe here the three methods I used when
building the various AM models. I
first make the spheroidal component (halo plus, when 
present, bulge) so that it is in equilibrium in its own potential plus
the monopole term of the disc potential (i.e. the spherically symmetric
equivalent of the disc). In the initial conditions built following
Hernquist (1993), this spheroid is simply stacked on the disc
component. Thus the simulation starts somewhat off equilibrium, but
settles after only a few transients. A second method consists in first
finding an equilibrium by running a constrained simulation in which
the spheroid and disc components are constrained to stay
axisymmetric. Once this is achieved, the full unconstrained simulation
can 
start with initial conditions either the previously found equilibrium,
or with disc velocities modified so as to follow given constraints,
e.g. a given profile of the $Q$ stability parameter (Toomre 1964). 
In the third method I grow the remaining terms (above the monopole)
of the disc
potential adiabatically in the spheroid, thus allowing it to settle to
a new equilibrium, as e.g. in  McMillan, Athanassoula \& Dehnen 
(2007) and in McMillan \& Dehnen (2007). I then remove the disc potential
and introduce the disc particles. In both the second and the third
method the simulation starts very near equilibrium. Obviously, in all
three methods, the initial conditions are not exactly as described by
Eqs.~\ref{eq:discdens}, \ref{eq:halodens} and
\ref{eq:hernquistp}, since the components have interacted between them
and adjusted to each other. Furthermore, the adjustments in the three
methods are
not identical, as expected. Nevertheless, and this is important to
stress, the simulations starting
from these three methods give the same global behaviour, evolution and
trends. More information on
how the disc component was set up can be found in Hernquist (1993),
in AM02 and in A03.   

I also included in this study three families of initial conditions with a
cuspy halo. In the first, the AM2 family, the initial halo profile is as in 
Eq.~\ref{eq:halodens}  with $\gamma$ of the order of
$10^{-2}$. In the second, the HRN family, the halo has initially a
Hernquist profile (Eq.~\ref{eq:hernquistp}). Finally in the third
family, the McM family, I used the initial conditions described by 
McMillan \& Dehnen (2007). 
For these simulations the halo has initially a profile:

\begin{equation} 
\rho_{\mathrm h} (r) = \frac{\rho_{\mathrm{c}}}{(r/r_{\mathrm{h}})^{\gamma_0}
(1 + r/r_{\mathrm{h}})^{3-\gamma_0}}   \textrm{sech}(r/r_{\mathrm{t}}),
\label{eq:tNFW}
\end{equation}

\noindent
where $\rho_{\mathrm{c}}$ is a density scale, $r_{\mathrm{h}}$ is the
halo scale radius, $r_{\mathrm{t}}$ is the halo truncation radius and
$\gamma_0$ measures the inner slope of the density profile.  
As in the previous families of models, the bulge here, whenever present, has
initially a Hernquist (1990) radial profile. The halo distribution
function is built using the Cuddeford (1991) inversion, while that of
the disc uses the method of Dehnen (1999).

I also ran simulations with the initial conditions of 
Kuijken \& Dubinski (1995), which constitute the KD family. In these
models, the bulge follows a King (1966) model and the halo has 
a lowered Evans distribution function
(Kuijken \& Dubinski 1994). The distribution function of the disc is a
three-dimensional generalisation of the planar distribution function of
Shu (1969) and Kuijken \& Tremaine (1992) and the radial density
profile on the equatorial plane is very near exponential. 
The Kuijken \& Dubinski (1995) model also allows one to
start with a flattened or with a rotating halo and I have used this in
a few cases.

In most of the simulations, all the particles
have the same mass. However, in a couple of dozen simulations of the
AM family the outer halo particles have a larger mass. As discussed in
A03, the largest masses are 
attributed to particles with the largest pericenters, thus ensuring
that these particles will not reach the inner regions. Of course the
halo distribution will change with time, but by being sufficiently
cautious about the pericenter limits, one can make sure that the high
mass particles will not reach the disc radii. In most
simulations of the McM family, I followed the precepts of McMillan \tal
(2007) and McMillan \& Dehnen (2007) and used halo particles with four
times the mass of the disc particles and halo softening twice as
large as that of the disc particles. Thus the
maximum force exerted by a single particle ($\propto m_i/\epsilon^2_i$, 
where $m_i$ and $\epsilon_i$ the mass and softening of the particle,
respectively) is the same for all particles. 

Some of these simulations were run on our 
GRAPE-5 systems (Kawai \tal 2000), while the others used
the public version of W. Dehnen's treecode (Dehnen 2000; 2002). As
already discussed in A03, the results obtained with the two codes are
in very good agreement. In most simulations the disc has 200\,000
particles and the halo about  
a million. I also ran several simulations with a considerably larger
number of particles. In all simulations the gravitational constant $G$
= 1.   

\begin{figure*}
  \setlength{\unitlength}{2cm}
  \includegraphics[scale=0.9,angle=-90]{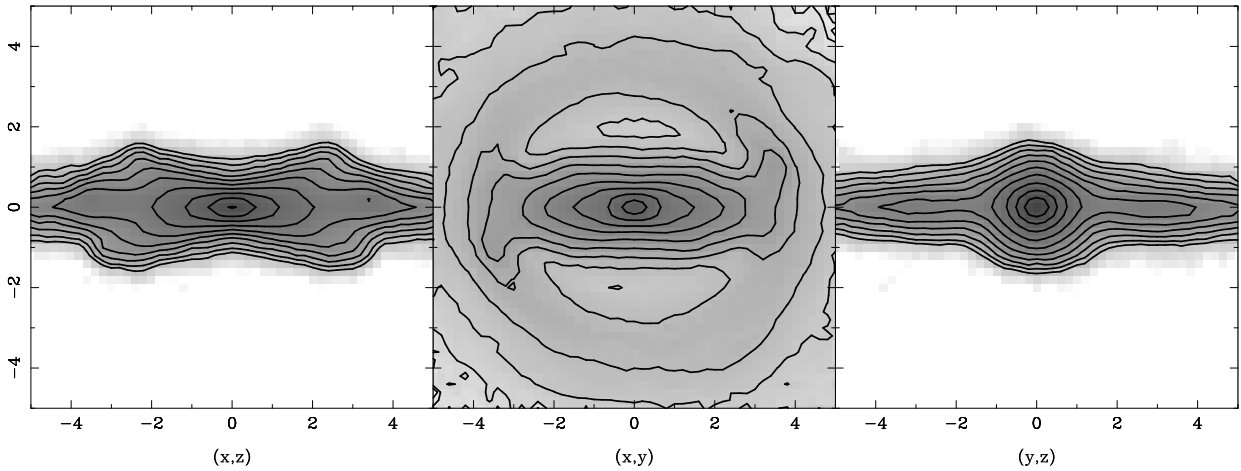}
\caption{Three orthogonal views of the disc component.
  The left panel gives the edge-on
  side-on view, the right one gives the edge-on end-on view and
  the middle one the face-on view. The projected density of the
  disc is given by 
  grey-scale and also by isocontours (spaced logarithmically). 
}
\label{fig:3disc}
\end{figure*}

\begin{figure*}
  \setlength{\unitlength}{2cm}
  \includegraphics[scale=0.9,angle=-90]{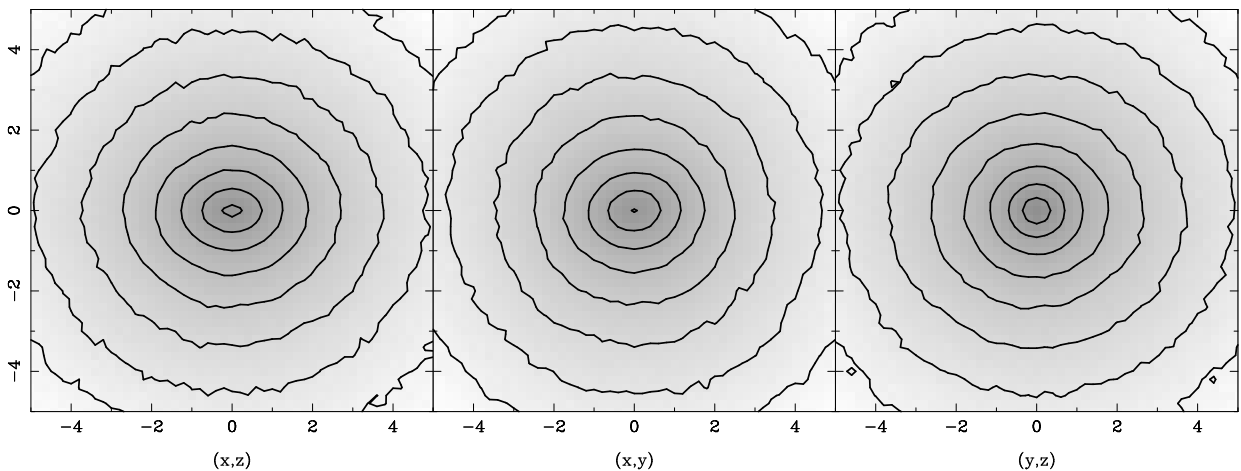}
\caption{As for Fig.~\ref{fig:3disc}, but for the halo component.
}
\label{fig:3halo}
\end{figure*}

The representative simulation discussed here
is part of the AM family with the non-monopole terms of the disc
potential grown adiabatically in the spheroid (last of the three
methods described 
above) and has $z_0$ = 0.2, $Q$ = 1.2, $M_h$ = 5, $\gamma$ = 0.5,
$r_c$ = 10 and  $M_b$ = 0, i.e. is of MH type. All plots and
results, unless otherwise stated, refer to time $t$ = 800, i.e. a time
towards the end of the simulations, well in the secular evolution
regime. Like in most simulations, the number of particles in the disc
is 200\,000 and about a million in the halo. This is sufficient for a
good description of the evolution, but can give relatively noisy
isodensity contours in a plot. In order to obtain smooth isocontours
without smoothing or loss of information, I used in
Figs.~\ref{fig:3disc} to \ref{fig:1halocut} the technique described in
Athanassoula (2005a). In Paper II, I will discuss 
statistics from all families of models presented here.

In order to convert computer units to units more readily used when
describing galaxies, I need to set the units of length and mass to
values representative of a barred
galaxy. AM02 used a length unit of 3.5 kpc and a mass unit of 
5 $\times$ $10^{10}$ $M_{\odot}$. This choice, however, is in no way
unique, and I can 
choose quite different values. Because of this,
I will not adopt here any particular calibration, but
stick to computer units. The reader can easily convert
units according to his/her needs and the objects under consideration.

\section{The halo bar}
\label{sec:existence}
\indent

Fig.~\ref{fig:3disc} shows the three orthogonal views of the disc
component of the representative 
simulation at time 800. Only the inner part is shown, in order to
show best the bar. Seen face-on, this has isodensities which are
elliptical-like in the inner parts and  rectangular-like further out.  
The bar is surrounded by a ring,
as the inner rings often seen in barred galaxies. Seen
side-on, the inner disc shows a strong peanut feature, again as often
observed. This is called by observers a peanut bulge. Seen
end-on, i.e. edge-on with the line of sight along the bar major
axis, the bar contributes a spherical-like central
feature. This has the same aspect as observed classical bulges in so called
non-barred galaxies (for a discussion of bulge types see
Athanassoula 2005a). 

Fig.~\ref{fig:3halo} shows the three corresponding orthogonal views of
the halo component. It is clear that the
halo distribution is not spherical. Particularly in the inner parts, it
has a prolate-like shape and seems to become less elongated at larger
radii. It also seems that the shortest axis is perpendicular to the
disc equatorial plane and the longest along the major axis of the disc
bar. For simplicity, I will call this elongated 
feature in the centre of the halo component the halo bar, although its
shape is more reminiscent of an oval. It is definitely less strong than the
disc bar, less eccentric and also shorter. Seen side-on it never has a peanut
shape. Its shape in all three projections looks roughly 
elliptical, so that the three-dimensional shape can be described as
ellipsoidal. It looks roughly aligned with
the disc bar. All these statements will be made more quantitative in
the next sections.

\begin{figure}
  \setlength{\unitlength}{2cm}
  \includegraphics[scale=0.45,angle=-90]{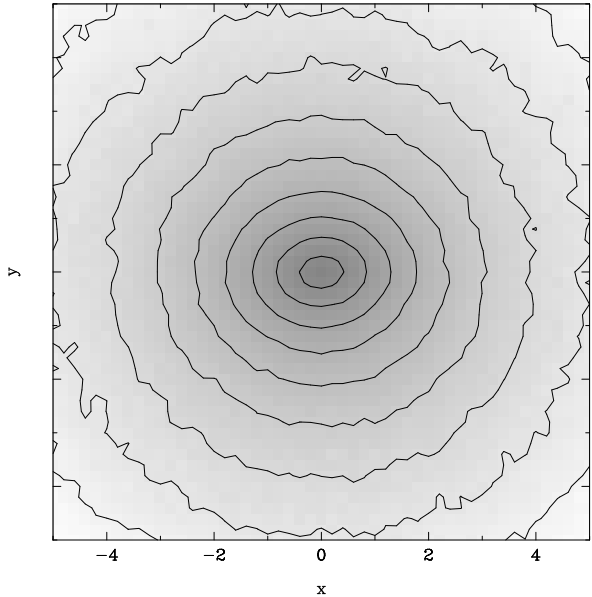}
\caption{$(x,y)$ view of the halo projected density. As for
  Fig.~\ref{fig:3halo}, but  
the projected surface density is now obtained only from
the mass within a slice around the disc equatorial plane,
i.e. is based only on particles with $|z|<1$. 
}
\label{fig:1halocut}
\end{figure}

Fig.~\ref{fig:3halo} has been made as Fig.~\ref{fig:3disc}, i.e. I
projected all the matter on the equatorial plane. However, since the
halo shape is triaxial in the inner parts and very near spherical in
the outer parts, this projection will make isodensities more round than
the true three dimensional shape. For this reason, I plot in
Fig.~\ref{fig:1halocut} again the isodensities, but this time taking
into account only particles with $|z| < 1$. A comparison of
Figs.~\ref{fig:3halo} and \ref{fig:1halocut} shows that the difference
is small. This is due to the fact that the halo is
fairly concentrated, so that the density in the outer parts is much
lower than in the inner parts.

\subsection{Halo inertia tensor and the axial ratio of its bar}
\label{sec:inert}
\indent

\begin{figure}
  \setlength{\unitlength}{2cm}
  \includegraphics[scale=0.35,angle=-90]{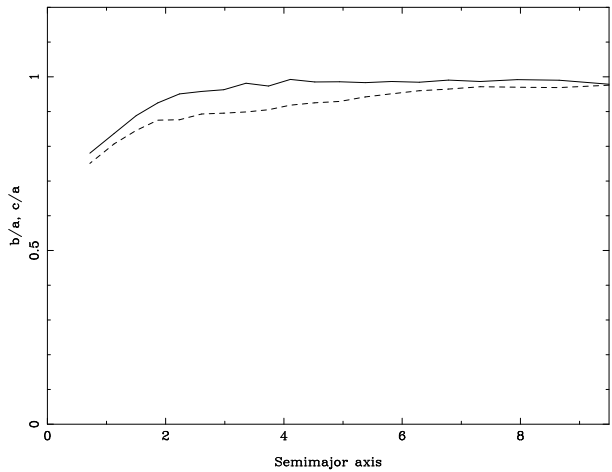}
\caption{Axial ratios as a function of semimajor axis for the halo
  mass distribution. The $b/a$ ratio is given
  with a full line and the $c/a$ by a dashed line.
}
\label{fig:axrat}
\end{figure}

Since the shape of the bar is roughly ellipsoidal, I used the
inertia tensor to calculate its axial ratio. For this I first
assigned a local density to each halo particle, by calculating the distance
to its nearest neighbours (Casertano \& Hut 1985) with the help of the
NEMO package (Teuben 1995). The particles were sorted in order of
increasing local density and divided in groups with local densities within a
given range, discarding those with the highest densities to ensure
that the distribution is not influenced by the softening. The axial
ratios for each group are obtained (e.g. Barnes 1992) as
those of the homogeneous ellipsoid that has the same moment of
inertia as the particle distribution. Tests showed that, if the sorting is
done with respect to the 
distance of the particles from the centre, then a bias towards
sphericity is introduced, since the shells into which the particles are
divided are necessarily spherical. A similar bias, although less strong, is
introduced if the particles are sorted by their potential or by their binding
energy, since the corresponding isocontours are more spherical-like
than the density ones. I adopted here a sorting by density since a
number of tests showed me that this does not introduce any bias and
gives very satisfactory results, its only disadvantage being that it
is rather CPU intensive.

I thus obtain
$
b/a = \sqrt{(q_2/q_1)}
$
and 
$
c/a = \sqrt{(q_3/q_1)},
$
where $q_1$, $q_2$ and $q_3$ are the eigenvalues of the inertia tensor,
$a$, $b$ and $c$ are the lengths of the three principal 
semi-axes of the halo and $a > b > c$. Their ratios are plotted in
Fig.~\ref{fig:axrat} as a function of the average semimajor axis of
the shell. The parts
within roughly four initial disc scale-lengths depart
substantially from sphericity, and more strongly so for regions nearer
to the centre. The innermost group of particles has axis ratios
somewhat less than 0.8. 
Within roughly the first 2 initial disc scale-lengths,
the two axial ratios do not differ substantially, i.e.
the general shape, although triaxial, is not far from
prolate. Somewhat further out and up to roughly 6 initial disc
scale-lengths $b/a$ is very near unity, while $c/a$ is smaller. Thus,
the shape is more oblate-like in this region. This is better
seen from measures of the triaxiality discussed below.  

\begin{figure}
  \setlength{\unitlength}{2cm}
  \includegraphics[scale=0.32,angle=-90]{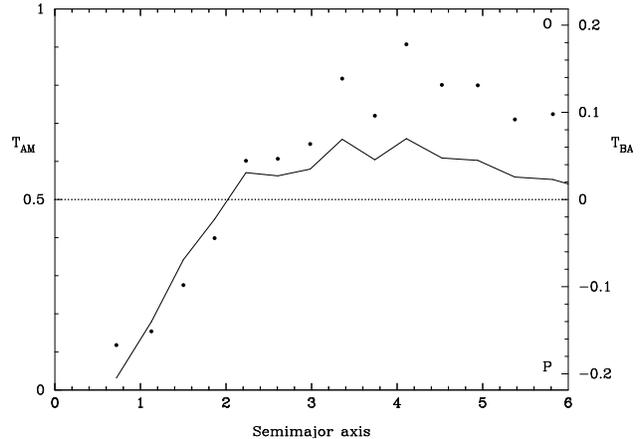}
\caption{Triaxiality parameters $T_{AM}$ (filled circles, ordinate to
  the left of the panel) and $T_{BA}$ (solid line, ordinate to the
  right of the panel) as a function of semimajor axis for the halo
  component. For both measures of triaxiality, the region 
  above the dotted horizontal line corresponds to oblate-like
  configurations and the region bellow it to prolate-like ones. 
}
\label{fig:triaxial}
\end{figure}

In Fig.~\ref{fig:triaxial}, I plot the triaxiality of the halo
component. Several definitions have been used so far. Aguilar \&
Merritt (1990) use  

\begin{equation}
T_{AM}=\frac {b-c} {a-c}.
\end{equation}

\noindent
$T_{AM}$ takes values in the range [0, 1], where the lower bound corresponds
to a prolate shape and the upper one to an oblate one. As
explained, however, in Boily \& Athanassoula (2006), $T_{AM}$
is sensitive to noise and to the corresponding small errors in the
calculation of the axes if 
$a\sim b\sim c$. To avoid this, Boily \& Athanassoula use   

\begin{equation}
T_{BA}={{b^2 - c^2} \over {b^2 + c^2}} - {{a^2 - b^2} \over {a^2 + b^2}}.
\end{equation}

\noindent
This takes
values in the range [-1, 1] and is positive for oblate shapes and
negative for prolate ones. 

The results of these two definition are compared in
Fig.~\ref{fig:triaxial}. Obviously, one should not compare the
numerical values given by the two definitions, since the two
mathematical forms are different. One should compare the general form
of the profile and the ranges where
the two say that the form is oblate-like, or prolate-like. Therefore,
Fig.~\ref{fig:triaxial} shows that the two definitions agree well. The
structure is prolate-like from the centre
to a semimajor 
axis $\sim$2. Beyond that and up to 6 the configuration is
roughly oblate-like, with a maximum oblateness for a semimajor axis
around 4. Beyond 6 the 
configuration is too near-spherical for any of the two methods to be
able to assess its shape (see also
Fig.~\ref{fig:axrat}).    

\subsection{Ellipse fitting and projected axial ratio of the halo bar}
\label{sec:ellipse}
\indent

\begin{figure}
  \setlength{\unitlength}{2cm}
  \includegraphics[scale=0.35,angle=-90]{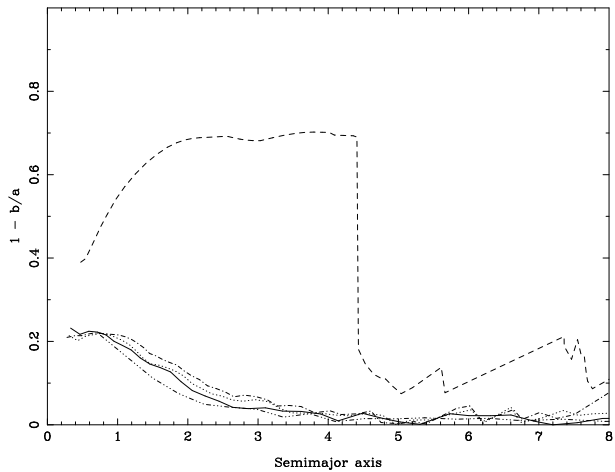}
\caption{Ellipticity in the equatorial plane of the bar in the disc 
  and the halo components. The dashed line gives results from ellipse
  fits to the disc density projected on the equatorial
  plane. The solid line corresponds to ellipses fitted similarly to
  the whole halo, the dot-dashed line to the halo truncated at $|z| <$
  1 and the dotted line to the halo truncated at $|z| < 2$. The
  dashed-triple-dot line gives the results from the halo inertia
  tensor. 
}
\label{fig:axratAM}
\end{figure}

As the shape of the disc bar is far from
ellipsoidal, AM02 did not use the
inertia tensor in order to measure the bar axial ratio in the
equatorial plane. Instead, they used the method introduced by
Athanassoula \tal (1990) for observations. 
Namely, they projected the density on the equatorial plane and 
fitted generalised ellipses to the isodensities. The equation of the
generalised ellipse, initially introduced by Athanassoula \tal (1990),
is

\begin{equation}
(|x|/a)^\lambda + (|y|/b)^\lambda = 1,
\end{equation}

\noindent
where $a$ and $b$ are the semimajor and semiminor axes, respectively,
and $\lambda$ is a parameter describing the shape of the generalised
ellipse. For $\lambda$=2 the shape is a standard ellipse,
for $\lambda<2$ it is a 
lozenge, while for $\lambda>2$ it approaches a rectangle, and, for
simplicity, is generally called rectangular-like. 

In Fig.~\ref{fig:axratAM}, I plot the ellipticity 1 - $b/a$ of the
disc and of the halo components, obtained as described above. The
profile for the disc resembles that shown in the upper left 
panel of Fig. 4 of AM02, as could be expected, since both results
are obtained for MH-type bars. It
has a relatively flat part and then drops abruptly at a semimajor
axis of $\sim$4.4. It thus provides one of the possible ways of
measuring the bar length for a strong bar (AM02). The halo bar
presents a totally different 
profile. The axial ratio increases slowly inwards either steadily, or
reaching at small radii a plateau of short extent. Contrary to the
disc bar, it has no sharp increase or decrease that could help  
determine the bar length. I can, nevertheless, distinguish two
separate parts to the profile, one in which the ellipticity decreases
with radius and the other where it is consistent with zero, to within
the noise. The transition occurs at a radius between 3 and
4.4 length units, i.e. within the disc bar length, but a sharper
evaluation is not possible. Furthermore, this length is not
necessarily equal to the halo bar length, but just constitutes an
upper limit. To define the halo bar length one needs to measure the
phase of the $m$ = 2 
component as will be done in $\S$~\ref{sec:length}. Here, I can
just conclude that the halo bar length can not exceed an upper
limit, estimated between 
3 and 4.4 length units, i.e. can not exceed the length of the disc bar. 
Fig.~\ref{fig:axratAM} also shows that the halo bar is much less
elongated than the disc 
bar, as could already be inferred by a comparison of
Figs.~\ref{fig:3disc} and \ref{fig:3halo}. E.g., at a distance of
two initial disc scale-lengths the ellipticity of the halo bar is
about seven times that of the disc bar.  

In the above, I treated the halo bar in exactly the same way as
the disc bar, i.e. I first projected the density on the disc
equatorial plane. Although this is a reasonable thing to do for 
the disc, it is less appropriate for
haloes, since these extend roughly the same in all 
directions and thus a projection on the
equatorial plane may artificially circularise the isodensities,
particularly in the inner regions. For this reason, I repeated
this ellipse fitting, but now truncating the halo so that particles with 
$|z|>1$, or $|z|>2$,  are ignored. The
results are also shown in Fig.~\ref{fig:axratAM}. Note that
the difference
is not large, so that one can get estimates of the halo bar 
axial ratio from any of these three measurements. As
already mentioned, this must be due
to the fact that the halo is centrally concentrated, so that its density
falls rapidly with distance from the centre.

In $\S$ \ref{sec:inert} I measured the halo bar ellipticity using the
inertia tensor of the three dimensional halo mass distribution. Although
this method is 
conceptually the most correct one, it has a clear drawback. Namely, a
relatively large number of particles are needed in order to estimate
the axial ratio of the inner parts and as a consequence the innermost
measurement can not be very near the centre. 
The ellipse fitting method discussed here does
not share this disadvantage, so it 
gives information for much smaller radii. Fig.~\ref{fig:axratAM}
compares the results of the two methods and
shows that they agree well, a welcome result,
since it allows the use of either method for statistical
measurements, such as will be given in Paper II. Note that the
innermost measurement from 
the inertial tensor is at $\sim 0.8$ length units, while the
innermost from the fitted generalised ellipses is at $\sim0.4$
length units.   

\section{Spherical harmonic analysis}
\label{sec:length}
\indent

\begin{figure}
  \setlength{\unitlength}{2cm}
  \includegraphics[scale=0.35,angle=-90]{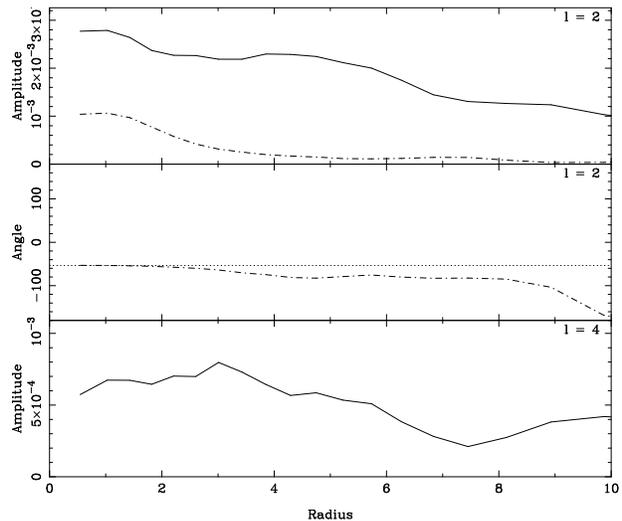}
\caption{Results of the spherical harmonic expansion of the halo mass.
  The upper panel gives the amplitude of
  the $l$=2 components. The $m$=0 component is given with a full
  line and the $m$=2 one with a dot-dashed line. The middle 
  panel gives the $l$=2 $m$=2 phase angle with a dot-dashed line and
  the position angle of the disc bar with a dotted line. The lower panel gives
  the amplitude of the $l$=4, $m$=0 component.
}
\label{fig:Ylm}
\end{figure}

In order to measure the strength of the disc bar it is customary, both
for observations and for simulations, to Fourier analyse the face-on
density distribution (e.g. Ohta, Hamabe \& Wakamatsu 1990; Ohta 1996;
AM02; Laurikainen \tal 2006; Buta \tal 2006). This method,
however, is not appropriate for the halo, which has a
three-dimensional density distribution. Instead, I expand the density,
$\rho(r, \theta, \phi)$, in spherical harmonics

\begin{equation}
\rho (r, \theta, \phi) = \sum_{l=0}^{\infty} \sum_{m=-l}^{l}
\rho_{lm} Y_l^m(\theta, \phi),
\end{equation}

\noindent
where $\phi$ is the azimuthal angle and $\theta$ is the meridional
angle, i.e measured
from the $z$ axis. $Y_l^m$ denotes the spherical harmonics defined as

\begin{eqnarray}
Y_l^m(\theta, \phi) = \sqrt {\frac{2l+1}
  {4\pi} \, \frac{(l-|m|)!}{(l+|m|)!}}~~~~~~~~~~~~~~~~~~~~~~~~~~~~~~~~~~~~~~~~~
  \nonumber  \\  
 \times \, P_l^{|m|}(\cos\theta) \, e^{im\phi}\left\{
  \begin{array}{ll}(-1)^m & \mbox{if $m \geq 0$}\\1 & \mbox{if $m <
  0$} \end{array} \right. 
\end{eqnarray}

\noindent
and

\begin{equation}
Y_l^{-m}(\theta, \phi) = (-1)^m Y_l^{-m*}(\theta, \phi),
\end{equation}

\noindent
so that

\begin{equation}
\rho_{lm} (r) = \int_0^\pi \int_0^{2\pi} d\phi \,
Y_l^{-m*}(\theta,\phi) \ 
\rho (r, \theta, \phi),
\label{eq:rholm}
\end{equation}

\noindent
where the * denotes a complex conjugate. 

The density of a sum of point masses can be written as 

\begin{equation}
\rho (r, \theta, \phi) = \sum_{k}~\frac{m_k}{r_k^2}\delta(r-r_k)
\delta(\phi-\phi_k)\delta(\cos\theta-\cos\theta_k),
\label{eq:rhoN}
\end{equation}

\noindent
where the index $k$ denotes the particle and $r_k$, $\phi_k$,
$\theta_k$ and $m_k$ are its radius, angle coordinates and mass,
respectively. The volume density, however, is not
well adapted to  $N$-bodies; instead it is more convenient to
work with the mass in a spherical shell. Using Eqs. \ref{eq:rholm} and
\ref{eq:rhoN} and integrating over the radius in a spherical
shell centred at radius $r$, I find

\begin{equation}
M_{lm} (r) = \sum_i m_i Y_l^{-m*}(\theta_i,\phi_i),
\label{eq:Mlm}
\end{equation}

\noindent
where the summation is carried over all particles in the shell
and where $r_i$, $\theta_i$ and
$\phi_i$ are the radius and angle coordinates of these particles. 

Expanding the volume density in spherical harmonics I can write

\begin{eqnarray}
\rho (r, \theta, \phi) =
\sum_{l=0}^{\infty}~\sum_{m=0}^{l}~P_{l}^{m}(\cos\theta)~~~~~~~~~~~~~~~~~~~~~~~~~~
 \nonumber \\
~~~~~~~~~~~~~~\times \left[\overline{A_{lm}}(r)~\cos(m\phi)~+
~\overline{B_{lm}}(r)~\sin(m\phi)\right],
\end{eqnarray}

\noindent
where I have combined terms with opposite sign of $m$, and where

\begin{equation}
\overline{A_{lm}}(r) =
N_{lm}\sum_{k}~\frac{m_k}{r_k^2}\delta(r-r_k)P_{l}^{m}(\cos\theta_k)
\cos(m\phi_k),
\label{eq:A}
\end{equation}

\begin{equation}
\overline{B_{lm}}(r) =
N_{lm}\sum_{k}~\frac{m_k}{r_k^2}\delta(r-r_k)P_{l}^{m}(\cos\theta_k)
\sin(m\phi_k),
\label{eq:B}
\end{equation}

\begin{equation}
N_{lm}=\frac{2l+1}{4\pi}(2-\delta_{m0})\frac{(l-m)!}{(l+m)!}
\end{equation}

\noindent
and $\delta_{m0}$=1 if $m$=0 and $\delta_{m0}$=0 otherwise. A similar
equation can be obtained for the mass within a thin spherical shell,
where the $\overline{A_{lm}}(r)$ and $\overline{B_{lm}}(r)$ are
replaced by 

\begin{equation}
A_{lm}(r) = N_{lm}\sum_{k}~m_k~P_{l}^{m}(\cos\theta_k)~\cos(m\phi_k),
\label{eq:Alm}
\end{equation}
 
\noindent
and

\begin{equation}
B_{lm}(r) = N_{lm}\sum_{k}~m_k~P_{l}^{m}(\cos\theta_k)~\sin(m\phi_k),
\label{eq:Blm}
\end{equation}

\noindent
and where, as in Eq. \ref{eq:Mlm}, the summation is carried over all 
particles in the shell and where $r$ is the mass-averaged mean radius of the
shell. The corresponding amplitudes are equal to 
$H_{lm} (r) = \sqrt {A_{lm}^2(r) + B_{lm}^2(r)}$
and can be used to measure the halo bar strength
(e.g. in Fig.~\ref{fig:Ylm} and \ref{fig:phasedif}). 

The amplitude of the various $l$=2 components is shown, after smoothing, in
Fig.~\ref{fig:Ylm}. The $l$=2, $m$=0 component is linked to the flattening of
the halo component towards the disc equatorial plane, while the $l$=2,
$m$=2 component is linked to the halo bar contribution, i.e. it would
have been zero for an axisymmetric system. It is clear that the $m$=2
component drops faster with distance from the centre and has roughly
disappeared at a radius of 4 initial disc scale-lengths,
while the $m$=0 component extends more than twice as far. This
explains the result found in $\S$~\ref{sec:ellipse}, namely that the
halo mass distribution is prolate in the inner parts and oblate in the
outer parts and also agrees
with the visual impression given by Figs.~\ref{fig:3halo} and
\ref{fig:axrat}, namely that the isodensities become axisymmetric
before becoming spherical. The $l$=0, $m$=4 stays
within the noise, showing that there is no quadrupole
component, in good agreement with the ellipse fitting results
described in the $\S$~\ref{sec:ellipse}. 

\begin{figure}
  \setlength{\unitlength}{2cm}
  \includegraphics[scale=0.9,angle=-90]{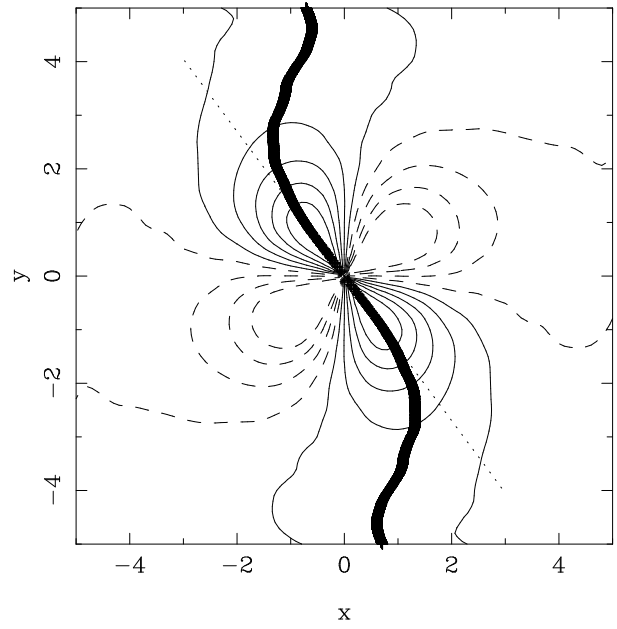}
\caption{Isocontours of the $l$=2 $m$=2 component of the halo mass
  distribution on the equatorial plane.
  Positive isocontours are given 
  with solid lines and negative ones with dashed lines. The thick
  line shows the phase of the halo bar and the thin dotted line gives
  the position angle of the disc bar.
}
\label{fig:isocl2m2}
\end{figure}

The middle panel of Fig.~\ref{fig:Ylm} compares the phase of the
$l$=2, $m$=2 component with that of the disc bar. A similar comparison
can also be made from Fig.~\ref{fig:isocl2m2}, which shows the
isocontours of this component on the equatorial plane. The two
figures, viewed together, show clearly that in the innermost parts the
phase of the halo bar does not change much with radius and that its value
is similar to that of the disc bar. At larger radii, however, the halo bar
phase clearly lags (trails) behind that of the disc bar, the difference
in angle increasing considerably with radius. At large distances, the 
$l$=$m$=2 component 
of the halo can be described as a very open trailing spiral. Thus, it
will exert a torque on the disc bar and pull it backwards. 

\begin{figure}
  \setlength{\unitlength}{2cm}
  \includegraphics[scale=0.45,angle=0]{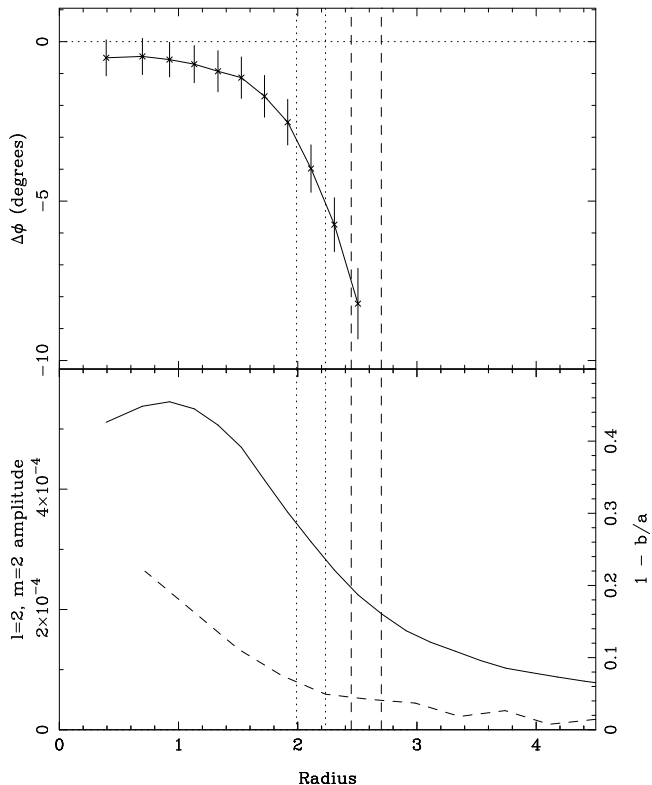}
\caption{Measuring the halo bar length. The upper panel shows the
  phase difference between the halo and the disc bar averaged between
  times $700 < t < 900$, as a 
  function of distance from the centre. The values
  represent averages over time and the vertical error bars have a
  length of one $\sigma$, as explained in the text. The vertical dotted
  and dashed lines give the one $\sigma$ range for the length of the
  halo bar (see text). The lower panel gives the radial profile of the $l$
  = 2, $m$ = 2 amplitude (solid line, ordinate to the left) and of the
  ellipticity (dashed line, ordinate to the right). The dashed and
  dotted vertical lines are continuations from the upper panel.   
}
\label{fig:phasedif}
\end{figure}

To look in more detail at the behaviour of the halo bar in the
innermost parts and in particular to get more information on its
phase relative to that of the disc bar, I need to average
over time, to improve the signal-to-noise ratio. Of course, this can
only be done towards the end of the simulation, when the time
evolution is slow. A typical result is shown in
Fig.~\ref{fig:phasedif}, where I have averaged the results of nine
equidistant times in the interval $700 < t < 900$. The error bars have
a length of one $\sigma$ and $\Delta \phi$ is measured from the phase of
the disc bar. Note that even the innermost point lags behind the disc
bar, but with a difference which is less than half a degree. This
difference increases with increasing distance from the centre. It is
thus not possible to say what the difference between the phases of the
disc and the halo bar is and this can well explain the differences
between the values quoted in the literature, which are thus not
necessarily due to differences between the simulations, but could
simply be due to the fact that the measurements were made at different
radii. The lengths of the error bars also increase with distance from
the centre, but remain reasonably small, at least for simulations
where the $l$ = $m$ = 2 component is sufficiently strong. 

Using the results of $\S$~\ref{sec:existence} and
\ref{sec:length}, it is possible to define the halo bar length in
three different ways. From the change of  
its phase, or from the radius at which the $l$=2, $m$=2 amplitude drops
to a certain fraction of its maximum, or from the ellipticity $1 -
b/a$ (see AM02 for corresponding 
definitions of the disc bar length). All three definitions include a
certain amount of arbitrariness, since in the first one needs to define what
the maximum allowed phase change is, and in the second one what fraction of
the maximum amplitude should be used. Finally, the third method would
not be arbitrary if, as in disc galaxies, the ellipticity profile had
a clear feature which one could associate to the bar length. This,
however, is not the case for the halo. To illustrate this
arbitrariness, I measure the length of the bar by the first definition
and a phase difference of 5$^\circ$ (or 10$^\circ$) and give the
results by vertical lines in the upper panel of 
Fig.~\ref{fig:phasedif} for the times used in 
that figure. The interval between the two dotted (dashed) lines
corresponds to one sigma of the bar length measurements 
and gives an estimate of the increase of
the halo bar length during that interval of the evolution. These lines
are continued in the lower panel, which shows the radial profile of
the $l$ = 2, $m$ = 2 spherical harmonic and of the
ellipticity. It is clear that the bar length when measured
by the phase shift will be shorter than when measured in the two
other ways, since there is a drastic change of the phase with radius
at radii at which the amplitude of the $l$ =2, $m$ = 2 component is
still quite high. In fact, measurements from the radial profile of
this amplitude, or from the ellipticity will give information on the
bar and spiral combined, i.e. give the radial extent of a feature
which is much larger than the bar. Thus, 
the correct way of measuring the halo bar length is from the phase
shift. It is interesting to note that this length, as measured by a
phase difference of 5$^\circ$, agrees well with the
location at which the triaxiality changes from prolate to oblate
(Fig.~\ref{fig:triaxial}). Since these measurements are made in totally
independent ways, this agreement gives confidence that the phase shift gives a
relatively accurate estimate of the halo bar length. It also 
shows clearly that the halo bar is much shorter than the disc bar and
that the measurements from the radial profile of the $l$ =2, $m$ = 2
component would have given an overestimate.  

Since the phase of the halo bar is roughly equal to that of the disc
bar at all times, the two bars must rotate with roughly the same
pattern speed and thus have roughly the same corotation radius. 
Furthermore, the halo bar is considerably shorter than the disc
bar, i.e. it is a slow bar with a corotation radius considerably
larger than the bar length. 
 
\section{Kinematics of the halo component}
\label{sec:kinematics}
\indent

Several $N$-body simulations have shown that the halo can absorb
angular momentum (e.g. Sellwood 1982; Debattista \& Sellwood 2000;
A02; A03; O'Neill \& Dubinski 2003; Valenzuela \& Klypin 2003; 
Martinez-Valpuesta \tal 2006). A03 found a correlation
between the angular momentum absorbed by the halo and the bar strength
(see Figs. 16 and 17 of that paper). One can thus expect some 
rotation in the halo component of barred galaxies, particularly if the
bar is strong, and I will show in this section that this is indeed the case. 

\begin{figure}
  \setlength{\unitlength}{2cm}
  \includegraphics[scale=0.45,angle=0]{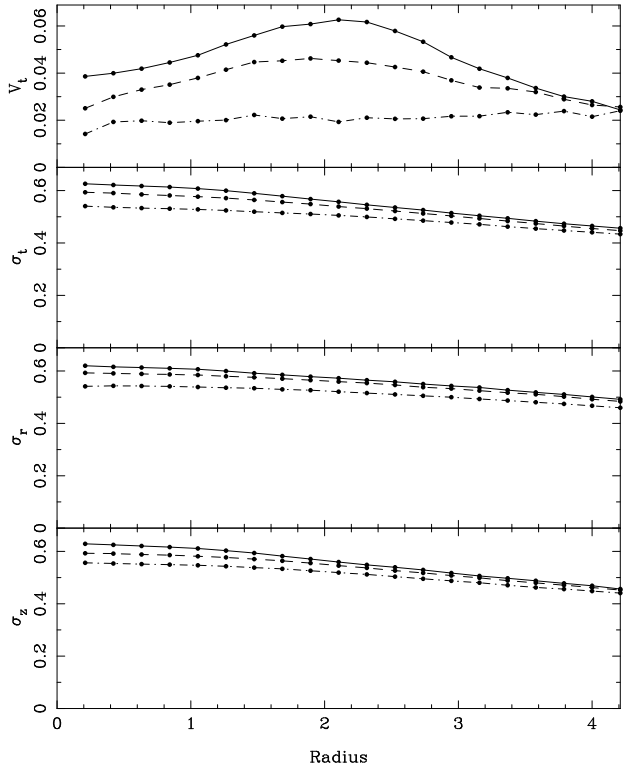}
  \caption{Radial profile of the tangential velocity (upper panel) and
  tangential 
  (second panel), radial (third panel) and $z$ component (lower panel)
  of the velocity dispersion, all as a function of
  cylindrical radius. Particles were
  split according to their $z$ coordinate. Results for particles very
  near the equatorial plane ($|z| < 0.5$) are given with a solid
  line, results for intermediate particles ($||z| - 1| < 0.5$) with
  a dashed line and results for particles further up ($||z| - 2| <
  0.5$) with a dot-dashed line.
}
\label{fig:Vsigma}
\end{figure}

Fig.~\ref{fig:Vsigma} gives information on the kinematics of the halo
component. The upper
panel shows the rotation curve (i.e. the average tangential velocity
for particles within a cylindrical annulus as a function of the radius
of the centre of the annulus) for three groups of particles :
particles with $|z| < 0.5$, particles with $0.5 < |z| < 1.5$ and
particles with $1.5 < |z| < 2.5$. Note that there is important
difference between the three, in the sense that particles near the
equatorial plane rotate considerably faster than particles further
away from it. The particles nearest to the
equatorial plane, rotate considerably, with a $V_{max}$ of 0.063
(compared to a $V_{max}$ of 0.77 for the disc component) and a 
$V_{max} / \sigma_{r} (R=0)$ of 0.11, where $\sigma_{r} (R=0)$ is the
radial velocity dispersion at the centre. This is well below the
oblate rotator line (Binney \& Tremaine 1987), which, for an axial
ratio of 0.8 and an isotropic velocity dispersion tensor, has a
$V/\sigma_{r}$ value $\sim$0.64. The rotation curve of this group of 
particles is not flat. It rises slowly, reaches a maximum at $R$ = 2.1
and then drops at a rate comparable to that at which it increases at
small radii. The maximum is located much further out than 
the maximum of the disc circular rotation curve, which
is at a radius less than one, and is
comparable to the halo bar length (see $\S$~\ref{sec:length}).   
Similar remarks can be made also for particles in the intermediate
$|z|$ bin.

\begin{figure}
  \setlength{\unitlength}{2cm}
  \includegraphics[scale=0.45,angle=0]{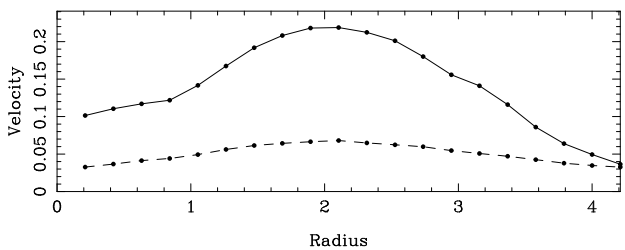}
  \caption{Tangential velocity as a function of cylindrical radius 
  at time 900. Only particles very
  near the equatorial plane ($|z| < 0.5$) are used. The solid line
  takes into account all particles that fulfil this criterion for
  $820 \le t \le 900$, while the dashed line takes into
  account all particles that fulfil this criterion at $t$ = 900.
}
\label{fig:Vres}
\end{figure}

The values of the rotational velocity are even bigger if, instead of using all
particles which at the time of the measurement have a $|z| < 0.5$, I
take all particles that fulfil this condition at all times in a given
time interval. This is clear from Fig.~\ref{fig:Vres}, which compares the
rotation curve from all particles that have $|z| < 0.5$ at $t$ = 900,
with the rotation curve obtained from particles that have $|z| <
0.5$ at all times $820 \le t \le 900$. Note that the latter has a
maximum velocity more than three times higher than the former. The reasons
for this are going to be discussed further in Paper II. It should be
noted, however, that even these 
particles have a considerably smaller rotational velocity than the
disc particles.   

The lower three panels of Fig.~\ref{fig:Vsigma} show the three
components of the velocity dispersion. All
three have a maximum at the centre, as expected, and
decrease with increasing radius. By comparing them, one can see
that the halo has a velocity distribution not far from isotropic.  The
three curves in each panel were 
obtained by binning particles by their $|z|$, as for the
tangential velocity (upper panel).
It is clear that particles with higher $|z|$ give larger
$\sigma$ values, but the difference is not as large as for the
tangential velocity.

\section{Summary}
\label{sec:summary}
\indent

A bar is not the monopoly of the disc component; such a structure is
also found in the halo of strongly barred galaxies. It has been called
the halo bar (here), or the dark matter bar (Colin \tal 2006). 
I make here a detailed analysis of the properties of a fiducial
model. As the disc bar, the halo bar can be found in the inner 
parts of the 
galaxy and it is roughly centred on the galaxy centre. Using the inertia
tensor, I calculated its axial ratios and found that it is triaxial
with the short axis along the axis of rotation. It is most elongated
at the innermost parts, with an axial ratio of the order of .8, and
turns gradually to spherical with 
increasing distance from the centre. Using two different measures of
the triaxiality, I show that the halo mass distribution is
prolate-like out to a certain radius and oblate-like beyond that. I also
looked at its shape using ellipse fits, in a way 
analogous to the disc bar, in order to make comparisons. I show that
the equatorial plane ellipticity decreases gradually with distance from
the centre, but has no sharp drop to provide a
measurement of the halo bar length. 

I expand the halo density in spherical harmonics and analyse the properties
of the $l$ = 2 and $l$ = 4 components. The $l$ = 2, $m$ =2 component
decays faster than the $l$ = 2, $m$ = 0 one, explaining the fact that
the halo mass distribution is prolate in the inner parts and oblate
further out. The 
amplitude of the $l$ = 4, $m$ = 0 component is considerably smaller
(about a factor of three or four) than that of the $l$ = 2, $m$ =
0. Spherical harmonics also provide a measurement of the halo
bar phase. I find that in the centermost part the halo bar lags the
disc bar by very 
little (of the order of a degree), but that the difference increases
with increasing distance from the centre, initially slower and then
considerably faster. It becomes larger than 5$^\circ$ at distances
somewhat larger than 2 initial disc scale-lengths. This, and in
particular the isocontours of the $l$ = 2, $m$ = 2 component, show that
the halo deformation is indeed bar-like in the inner parts and becomes
spiral-like further out, always lagging behind the disc bar. I use the
phase of the $l$ = 2, $m$ =2 component to measure the bar
length, and find that the end of the bar roughly coincides with the
position where the triaxiality, as measured from the inertia tensor,
turns from prolate to oblate. Since these two measurements are done in
a completely independent way, this agreement gives me confidence that
I am measuring correctly the halo bar length. The amplitude of the $l$
= 2, $m$ =2 component is large even beyond the end of the bar and this
can be understood by the fact that beyond the bar there is a spiral of
considerable amplitude.

I also analyse the kinematics of the halo component and find that
there is indeed some rotation, although it is largely constrained to
a layer around the equatorial plane. The maximum of the rotation
occurs not far from the end of the halo bar. The velocity
dispersion is not far from isotropic. It also
decreases with distance from the equatorial plane, but the effect is
much less than for the tangential velocity. The $V/\sigma$
value shows that the halo is a slow rotator. 

A general discussion of the implications of the results presented in
this paper will come after Paper II. In that paper I will apply the
techniques introduced 
here to a large sample of several hundred simulations, with different
initial conditions described in $\S$~\ref{sec:simul}, and I will
make comparisons and statistics. I will also discuss the time
evolution of the various halo bar properties and parameters.
 
There is a general qualitative agreement between our results and those
presented 
in previous studies mentioned in the Introduction. The comparison,
however, can only be fragmentary, since there is no previous complete
study covering all the aspects of the halo properties considered here.   
Furthermore, a
quantitative comparison is not possible at this stage, since, as I
will show in Paper II, there are considerable differences from one model
to the other and many properties show trends as a function of the
model parameters. 

\section*{Acknowledgements}

I thank A. Bosma, A. Misiriotis and K. Holley-Bockelmann for
many useful discussions and/or email exchanges and Jean-Charles
Lambert for computing assistance.
I also thank the INSU/CNRS, the region PACA 
and the University of Aix-Marseille I for funds to develop the
computing facilities used for the simulations discussed in this
series of paper. This work started while I was visiting INAOE. I would
like to thank ECOS-Nora and ANUIES for
financing this trip and my collaborators in INAOE for their kind hospitality.

\bibliographystyle{mn2e}

\bibliography{halobar_propI}

\label{lastpage}

\end{document}